\documentclass{appolb} 
\usepackage{epsfig} 
\usepackage{amssymb} 
\usepackage{color} 

\begin{document} 

\title{Asymmetric random matrices: what do we need them for?} 

\author{Stanis{\l}aw Dro\.zd\.z$^{1,2}$, Jaros{\l}aw Kwapie\'n$^1$, Andreas A.~Ioannides$^3$ 
\address{$^1$Institute of Nuclear Physics, Polish Academy of Sciences, PL--31-342 Krak\'ow, Poland \\ 
$^2$Faculty of Mathematics and Natural Sciences, University of Rzesz\'ow, PL--35-310, Rzesz\'ow, Poland \\ 
$^3$Laboratory for Human Brain Dynamics, AAI Scientific Cultural Services Ltd., 33 Arch. Makarios III Avenue, Nicosia 1065, Cyprus}} 

\maketitle 

\begin{abstract} 

Complex systems are typically represented by large ensembles of observations. Correlation matrices provide an efficient formal framework to extract information from such multivariate ensembles and identify in a quantifiable way patterns of activity that are reproducible with statistically significant frequency compared to a reference chance probability, usually provided by random matrices as fundamental reference. The character of the problem and especially the symmetries involved must guide the choice of random matrices to be used for the definition of a baseline reference. For standard correlation matrices this is the Wishart ensemble of symmetric random matrices. The real world complexity however often shows asymmetric information flows and therefore more general correlation matrices are required to adequately capture the asymmetry. Here we first summarize the relevant theoretical concepts. We then present some examples of human brain activity where asymmetric time-lagged correlations are evident and hence highlight the need for further theoretical developments. 

\end{abstract} 

\section{Complexity and matrices} 

One of the central concepts in contemporary science is complexity. In qualitative terms this concept refers to diversity of forms, to emergence of coherent patterns out of randomness and, at the same time, to an often observed impressive ability of switching among such patterns. In most cases approaching complex systems, either empirically or theoretically, is based on analyzing large multivariate ensembles of parameters. Therefore, an efficient formal frame to quantify various effects connected with complexity is in terms of matrices~\cite{Drozdz2002}. Since complexity primarily involves chaos, or even noise, the random matrix theory (RMT)~\cite{Mehta1991,Guhr1998} provides an appropriate reference. The RMT results provide a reference for quantification of the generic properties of a system that can provide clues of the underlying structure and its relation to chaotic or noisy nature of its dynamics. The system components that correspond to statistically significant excursions from the RMT distribution constitute the essence of complexity. These components reflect a creative and thus also deterministic potential emerging from an overwhelming noisy background in such systems~\cite{Drozdz2001}. Of great relevance in this context are correlation matrices that represent multivariate empirical observations or cases~\cite{Muirhead1982}. Up to now most of the practical implementations of such matrices deals with the symmetric cases (symmetric correlation matrices) which in the limit of purely random correlations corresponds to the Wishart ensemble~\cite{Wishart1928} with the corresponding eigenvalue distribution as represented by the Mar\u cenko-Pastur formula~\cite{Marcenko1967}. Complexity in real world systems often involves asymmetry in long-distance interactions and correlations both in space and in time. For instance, the information flow takes time, especially on the longer distances, which results in time-delayed correlations. The first in the literature documented attempts to handle such effects with the use of asymmetric correlation matrices deals with correlations in the human brain~\cite{Kwapien2000} and in the financial markets~\cite{Kwapien2006,Biely2008}. We present below new analysis of brain activity that reveals subtle asymmetric effects that identify the pressing need to extend the formalism of asymmetric random matrices and point to the direction such research should follow.

\section{Asymmetric correlation matrix}

Standard correlation matrix analysis for a system with $N$ degrees of freedom can straightforwardly be generalized to the case of two separate systems $\Omega_1,\Omega_2$ with $N$ degrees of freedom each. Let then the observable $X_{\alpha}$ account for each of the degree of freedom $\alpha$ of the system $\Omega_1$ and the observable $Y_{\beta}$ for each of the degrees of freedom of the system $\Omega_2$. Correspondingly, let $\{ x_{\alpha}(t_i) \}$ and $\{ y_{\beta}(t_i) \}$ denote the time series of the related measurements at $i=1,...,T$. In order to allow a full generality, the time series representing the system  $\Omega_2$ can be considered shifted in time by an interval $\tau = m \Delta t$ ($m$ is an integer number) with respect to their $\Omega_1$ counterparts. One then considers two data matrices:
\begin{equation} 
X_{\alpha,i} = {1 \over \sigma_{\alpha}} (x_{\alpha}(t_i) - \bar{x}_{\alpha}) 
\qquad Y_{\beta,i}(\tau) = {1 \over \sigma_{\beta}} (y_{\beta}(t_i+\tau) - 
\bar{y}_{\beta}), 
\end{equation} 
which can be used to form a general asymmetric correlation matrix 
\begin{equation} 
{\bf C}(\tau) = {1 \over T} {\bf X} [{\bf Y}(\tau)]^{\rm T}. 
\label{eqn::correlation.matrix.asymmetric} 
\end{equation} 
of the size $N \times N$. Its diagonal matrix elements no longer have to equal 
unity and thus ${\rm Tr} 
{\bf C} \le N$. Determination of the corresponding eigenvalues and 
eigenvectors demands solving the $\tau$-dependent secular equation: 
\begin{equation} 
{\bf C}(\tau) {\bf v}^{(k)}(\tau) = \lambda_k(\tau) {\bf v}^{(k)}(\tau). 
\end{equation} 
In general the matrix ${\bf C}(\tau)$ is asymmetric and thus the eigenvalues 
$\lambda_k(\tau)$ and the expansion coefficients $v_{\gamma}^{(k)}(\tau)$ are 
complex. This matrix remains however real therefore the eigenvalues and the 
expansion coefficients form the complex conjugate pairs. The real part of the 
spectrum is related to the symmetric component of the matrix ${\bf C}(\tau)$ 
and the imaginary part of the spectrum to its asymmetric component. Ordering 
of eigenvalues is determined by the condition: $|\lambda_k| \ge 
|\lambda_{k+1}|$, with the supplementary condition ${\rm Im} \lambda_k > {\rm 
Im} \lambda_{k+1}$ for a pair of the complex conjugate values. 

\section{Random matrix reference} 

A standard way to identify the real information content of the correlation matrix is to test it against a null hypothesis of completely random correlations characteristic for independent signals. The most appropriate ensemble of random matrices that can serve as reference for the above asymmetric correlation matrices is generated by products of two different rectangular $N \times T$ matrices (counterparts of ${\bf X}$ i ${\bf Y}$) with the Gaussian distribution of elements. Up to now there exists no derivation of the analytically closed formula describing distribution of eigenvalues as a function of $Q=T/N$ for such an ensemble. Some preliminary investigations in this direction~\cite{Biely2008} indicate a characteristic enhanced density of eigenvalues along the real axis and for the eigenvalues that are dispersed on the complex plane their 'clustering' around the origin. This 'clustering' gets however dissolved for $N \to \infty$ and $T \to \infty$ such that $Q$ remains a constant. Similar effect of clustering and its asymptotic disappearance is also observed~\cite{Kanzieper2010} for the complex valued correlation matrices. In the later case however no enhanced density of eigenvalues along the real axis takes place. Temporarily thus, for sufficiently large values of $N$ and $T$ parameters, the real asymmetric correlation matrices as defined by Eq.~\ref{eqn::correlation.matrix.asymmetric} can be considered to have properties deviating least from those of the orthogonal Ginibre ensemble (GinOE) of asymmetric random matrices~\cite{Ginibre1965} for which the spectral distribution of eigenvalues is known analytically. For this reason the Ginibre ensemble can be used as a first approximation reference for the empirical matrices (\ref{eqn::correlation.matrix.asymmetric}). 

GinOE matrices ${\bf G}$, being generalization of the GOE matrices, are defined by the Gaussian distribution of elements:
\begin{equation} 
p({\bf G}) = (2 \pi)^{-N^2/2} \exp [-{\rm Tr} ({\bf G}{\bf G}^{\rm T})], 
\end{equation} 
where ${\bf G}$ is of the $N \times N$ dimension, and a variance of the distribution $\sigma^2 = 1$. Spectrum of eigenvalues of such matrices decomposes itself into $N-L$ complex values and $L$ real values where the expectation value of $L$ asymptotically behaves as~\cite{Edelman1994}:
\begin{equation} 
\lim_{N \to \infty} E(L) = \sqrt {2 N \over \pi}. 
\end{equation} 
For finite values of $N$ it can be approximated by: 
\begin{equation} 
E(L) = 1/2 + \sqrt {2 N \over \pi} \left( 1 - {3 \over 8 N} - 
{3 \over 128 N^2} + \mathcal{O}(N^{-3}) \right). 
\label{eqn::ginibre.real.expected} 
\end{equation} 
The distribution of eigenvalues $\lambda=\lambda_x + i \lambda_y$ on the complex plane is described by the following expression~\cite{Edelman1997,Sommers2008}:
\begin{equation} 
\rho_G(\lambda) = \rho_G^c(\lambda) + \delta (\lambda_y) \ \rho_G^r(\lambda), 
\label{eqn::ginibre.eigenval} 
\end{equation} 
in which: 
\begin{equation} 
\rho_G^c(\lambda) = {2 |\lambda_y| \over \sqrt{2 \pi}} \left( 1 - {\rm erf} 
(\sqrt{2} |\lambda_y|) \right) \  e^{2 \lambda_y^2} \int_{|\lambda_x|^2}^ 
{\infty} du \ e^{-u} {u^{N-2} \over \Gamma (N-1)}, 
\label{eqn::ginibre.complex} 
\end{equation} 
\begin{eqnarray} 
\rho_G^r (\lambda) &=& {1 \over \sqrt{2 \pi}} \int_{|\lambda_x|^2}^{\infty} du 
\ e^{-u} {u^{N-2} \over \Gamma(N-1)} \\ 
\nonumber 
&+& {1 \over \sqrt{2 \pi}} \ |\lambda_x|^{N-1} \ e^{-\lambda_x^2/2} \int_0^ 
{\lambda_x} du \ e^{-u^2/2} {u^{N-2} \over \Gamma(N-1)}. 
\label{eqn::ginibre.real} 
\end{eqnarray} 
The function ${\rm erf}(x)$ in Eq.~(\ref{eqn::ginibre.complex}) denotes the Gaussian error function. In the limit $N \to \infty$ the above expression simplifies such that the $\lambda$ values form a uniform circle on the complex plane and a uniform interval on the real axis~\cite{Sommers1988,Kanzieper2005,Forrester2007}: 
\begin{eqnarray} 
\rho_G^c (\lambda) = {1 \over \pi} \Theta (\sqrt{N} - |\lambda|),\\ 
\rho_G^r (\lambda) = {1 \over \sqrt{2 \pi}} \Theta (\sqrt{N} - |\lambda_x|), 
\end{eqnarray} 
where $\Theta(\cdot)$ denotes a Heaviside function. 

\begin{figure}
\hspace{1.8cm} 
\epsfxsize 8cm 
\epsffile{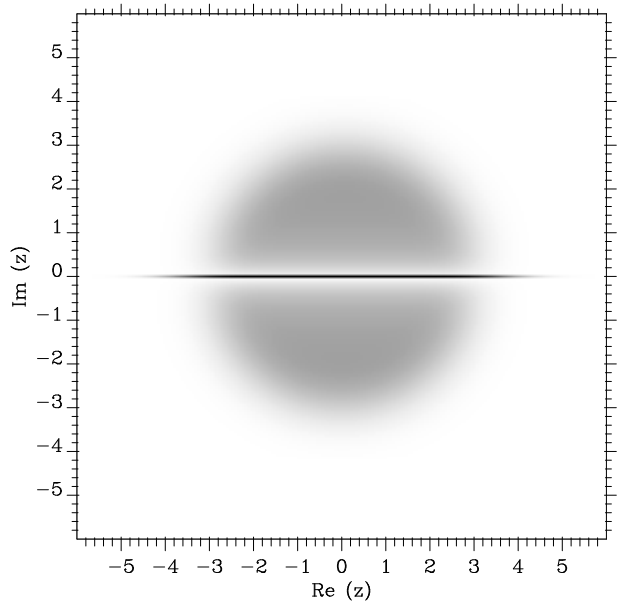} 
\caption{Theoretical eigenvalue distribution of $10 \times 10$ GinOE matrices in the complex plane. (\ref{eqn::ginibre.eigenval}). Darker regions correspond to higher eigenvalue density $\rho_G(z)$.} 
\label{fig::ginibre} 
\end{figure} 

One numerically generated example of the eigenvalue distribution of the random matrix ${\bf G}$ with the structure defined by Eq.~(\ref{eqn::correlation.matrix.asymmetric}) is shown in Fig.~1 for $N=10$ drawn from $10^6$ samples. Both, the circular shape of this distribution with its monotonically decreasing radial component $\rho_G(r)$~\cite{Biely2008} as well as a strip of enhanced density along the real axis are  consistent with the Eqs.~(\ref{eqn::ginibre.complex}) and (\ref{eqn::ginibre.real}). The results obtained from the empirical correlation matrix can be compared to such a GinOE ensemble after scaling $\lambda \mapsto \lambda / \sigma$, where $\sigma$ denotes the mean standard deviation of the distribution of matrix elements of ${\bf C}$ (if this distribution does not differ much from a Gaussian.

\section{Time-lagged correlations in the visual cortex}

Information transfer within the brain is associated with weak electric currents which generate an electric potential and magnetic field. When such currents in many nearby neuronal cells act coherently, the potential and fields grow large enough to be detected outside the skull. The corresponding techniques are known as Electroencephalography (EEG) and Magnetoencephalography (MEG), respectively. MEG~\cite{Ioannides2006} is particularly appropriate for studying the spatiotemporal patterns activity within the brain, including high-frequency ones; MEG has the same temporal resolution as the more conventional EEG allowing monitoring of neuronal activity down to the scale of 1 ms~\cite{Ioannides2006}. MEG and EEG are completely non-invasive methods of measuring the distribution and time dependence of the electric and magnetic fields outside the skull. Furthermore, the main advantage of MEG over scalp-EEG is that the skull and the scalp are transparent to the magnetic field and, therefore, an external measured magnetic field is only minimally distorted by the resistivity profile between the generators and sensors. In addition, the magnetic fields outside the skull are generated predominantly by the currents tangential to the surface of the head. The cortical currents are perpendicular to the surface of the cortex but almost $70\%$ of the human cortex is folded into fissures which makes these currents effectively tangential to the skull and, thus, accessible to MEG. The above aspects of MEG make it particularly attractive for studying the high-frequency spatiotemporal characteristics of the brain dynamics especially with the modern helmet-shaped probes~\cite{Kwapien1998}. 

Here, we extend our previous asymmetric correlation matrix analysis~\cite{Kwapien2000} of the time-lagged correlation between the left and right auditory cortices to a more challenging case of processing complex visual stimuli. Specifically, we present exemplary results from the analysis of MEG data recorded while a subject performed a visual object recognition task. In the experiment 30 different images were used, for each one of 5 categories (horses, trucks, birds, chairs, flowers). Each image was displayed for 0.5 s in front of a subject wearing an MEG helmet. His task was to recognize the seen object and select the proper name from a list presented to him a second later. During the whole presentation the magnetic activity of the subject's cortex was recorded with 510 Hz frequency by a 148-channel MEG apparatus (more technical details can be found in ref.~\cite{Ioannides2000}). Trials in which the displayed object was incorrectly named were removed, so out of all the 150 trials there were 140 trials left for further analysis.
 
Numerous Positron Emission Tomography (PET) and functional Magnetic Resonance Imaging (fMRI) studies have identified the relevant regions of activation in similar experiments. Out of many such regions, for further analysis we select three pairs of homologous areas in each hemisphere that showed prominent activations in our experiment and corresponded to areas identified by PET and fMRI. These are: posterior calcarine sulcus (PCS), fusiform gyrus (FG) and the amygdaloid complex (AM). PCS is located in the medial part of the occipital lobe and it takes a vital part in low-level processing of visual stimuli. It comprises both V1 and V2 areas that are the first cortical areas in the visual hierarchy. The FG, located between the temporal and occipital lobes, is involved in processing and selectivity of object images within specific categories. AM is an almond-shape structure located in the medial temporal lobe which is known to be crucial in the processing of emotions, especially fear, and in the recognition of emotional expressions in human faces. The PCS and FG are expected to be strongly involved in the processing of the stimuli used in the experiment; we expect also that the activity in the PCS and FG to be statistically correlated in each trial because the processing of stimuli requires exchange of information between such specialized areas. In contrast the AM is not expected to be particularly involved in the processing of the emotionally neutral objects used in the experiment. We will nevertheless investigate the temporal correlations in the activity from these regions of interest (ROIs) in the two hemispheres to test whether or not the above expectations are confirmed. 

\begin{figure} 
\hspace{1.8cm} 
\epsfxsize 8cm 
\epsffile{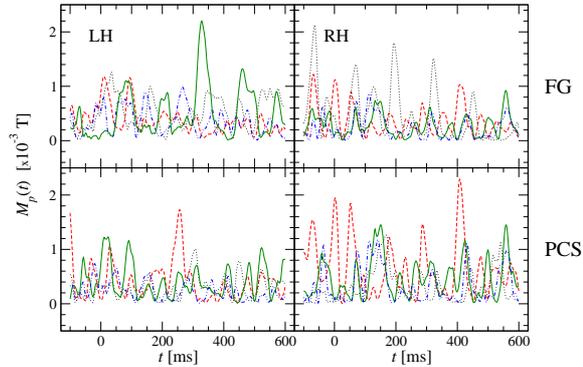} 
\caption{Exemplary single-trial signals analyzed in this work. The top row corresponds to fusiform gyrus and the bottom row to posterior calcarine sulcus, while the columns refer to the left and right hemispheres. Stimuli were presented from 0 to 500 ms. There is no evident stimulus-evoked activations of the regions.} 
\label{fig::signals}
\end{figure}

In order to extract a record of activity of the selected regions, a rather sophisticated and time-demanding procedure known as Magnetic Field Tomography (MFT) was applied~\cite{Ioannides1990,Taylor1999}. Output of this procedure were $N=140$ signals describing the total activity of the investigated regions during single trials. For the purposes of the illustrative examples of this paper we shall not distinguish between the categories presented in each trial in order to have better statistics, so we will use all (140) available clean trials. Each signal started 100 ms before the image presentation and ended 100 ms after the image had been switched off, therefore it covered 700 ms (the signal's length was $T=357$ data points, implying $Q=2,55$). A few exemplary signals are shown in Figure~\ref{fig::signals}. After preparing the signals we employed them to construct a family of $\tau$-lagged correlation matrices of size $N \times N$. For a given value of $\tau$ from the range $-50 \le \tau \le 50$, each signal representing ROI$_1$ was projected on each $\tau$-lagged signal representing ROI$_2$, so the resulting matrices were asymmetric (Eq.~(\ref{eqn::correlation.matrix.asymmetric})). We note that by considering the stimulus-locked time axis, we emphasize the activity evoked by the stimulus and de-emphasize the spontaneous activity of the ROIs. The matrices were then diagonalized and the complex eigenvalue spectra were derived for each value of $\tau$ ($\tau > 0$ denotes retarding of the second region in a pair, while $\tau < 0$ denotes the opposite). Figure 3 shows real part of the $\tau$-dependent largest eigenvalue $\lambda_1(\tau)$ for four different pairs of ROIs. Six pairs are unilateral (both the homologous pairs are considered in each case): PCS-FG (a), PCS-AM (b), FG-AM (c), and the last two pairs shown are formed across the hemispheres: PCS$^{(LH)}$-PCS$^{(RH)}$ and FG$^{(LH)}$-FG$^{(RH)}$ (d). As it is evident from Figure 3, $\lambda_1(\tau)$ for both the homologous PCS-FG pairs behaves distinctly from its counterpart for the other pairs.  For the PCS-FG pair a broad excursion above the threshold of $\rho_G^r(\lambda) = 0.01 \rho_G^r(0)$ (dashed horizontal line) is observed for positive and negative lags up to 70 ms. The pattern is sharper around the peak that is very nearly at the zero-lag origin for both the left and right hemispheres. This means that across single trials the activity in OCS and FG ROIs is strongly correlated with bi-directional flow of information in each hemisphere.  The flow of information is evenly distributed in each direction over the time range of latencies considered (-100 to 600 ms from stimulus onset). It is noteworthy that the correlations exceed the GinOE threshold (denoted by horizontal belts in Figure, $Q=2.55$) over a significant range of $\tau$ which can be explained by contributions from a range of frequencies that include the low-frequency oscillation observed in the signals (Figure 2). On the other hand, no such prominent maximum is seen in other pairs of ROIs; they hardly exceed the RMT threshold for PCS-AM and FG-AM. This suggests that during the experiment AM had a drastically different pattern of activity not related to the activation of the visual cortex (as expected). Yet another situation can be found for the pairs of homologous regions (Figure 3(d)): there are values of $\tau$ which seem to be statistically significant (especially for -70 ms $< \tau <$ 20 ms) but they are of a rather moderate magnitude. This result suggests that the activity in each FG ROI is linked to that of the homologous FG ROI in the other hemisphere, but the link is not as strong as that with the PCS on the same side. In the context of Figure 3(b) and 3(c), a question emerges to what extent the maximum values of $\lambda_1(\tau)$ can be considered non-random if they are placed near (slightly below or slightly above) the GinOE threshold. It should be recalled from the previous section, however, that the GinOE is not a fully relevant matrix ensemble for being a reference for the asymmetric correlation matrices, so the thresholds used in Figure 3 might not be exactly appropriate. This problem indicates the urgent need for developing the exact analytic results for the asymmetric Wishart ensemble of random matrices. 

\begin{figure} 
\hspace{1.8cm} 
\epsfxsize 8cm 
\epsffile{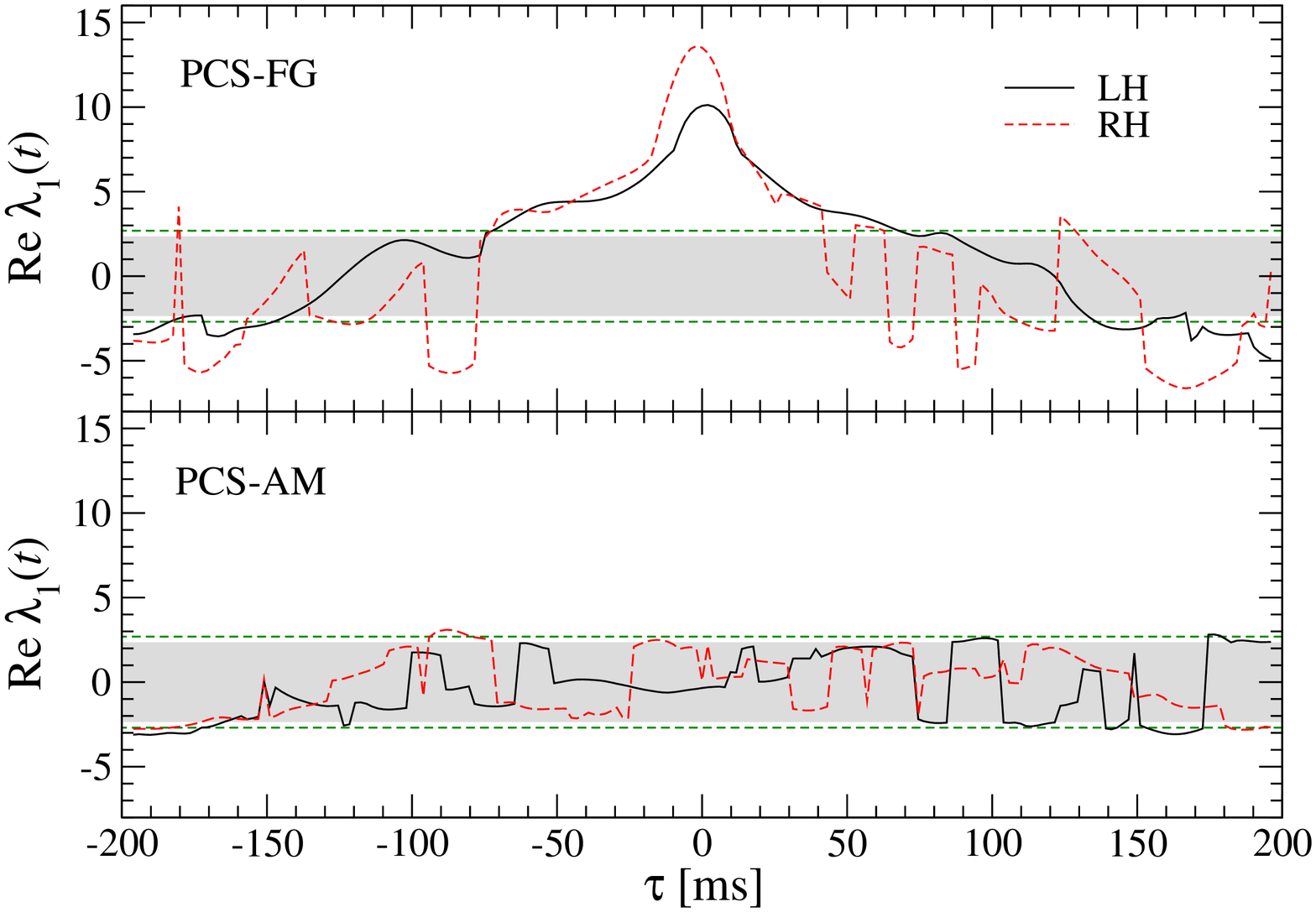} 

\vspace{-0.5cm} 
\hspace{1.8cm} 
\epsfxsize 8cm 
\epsffile{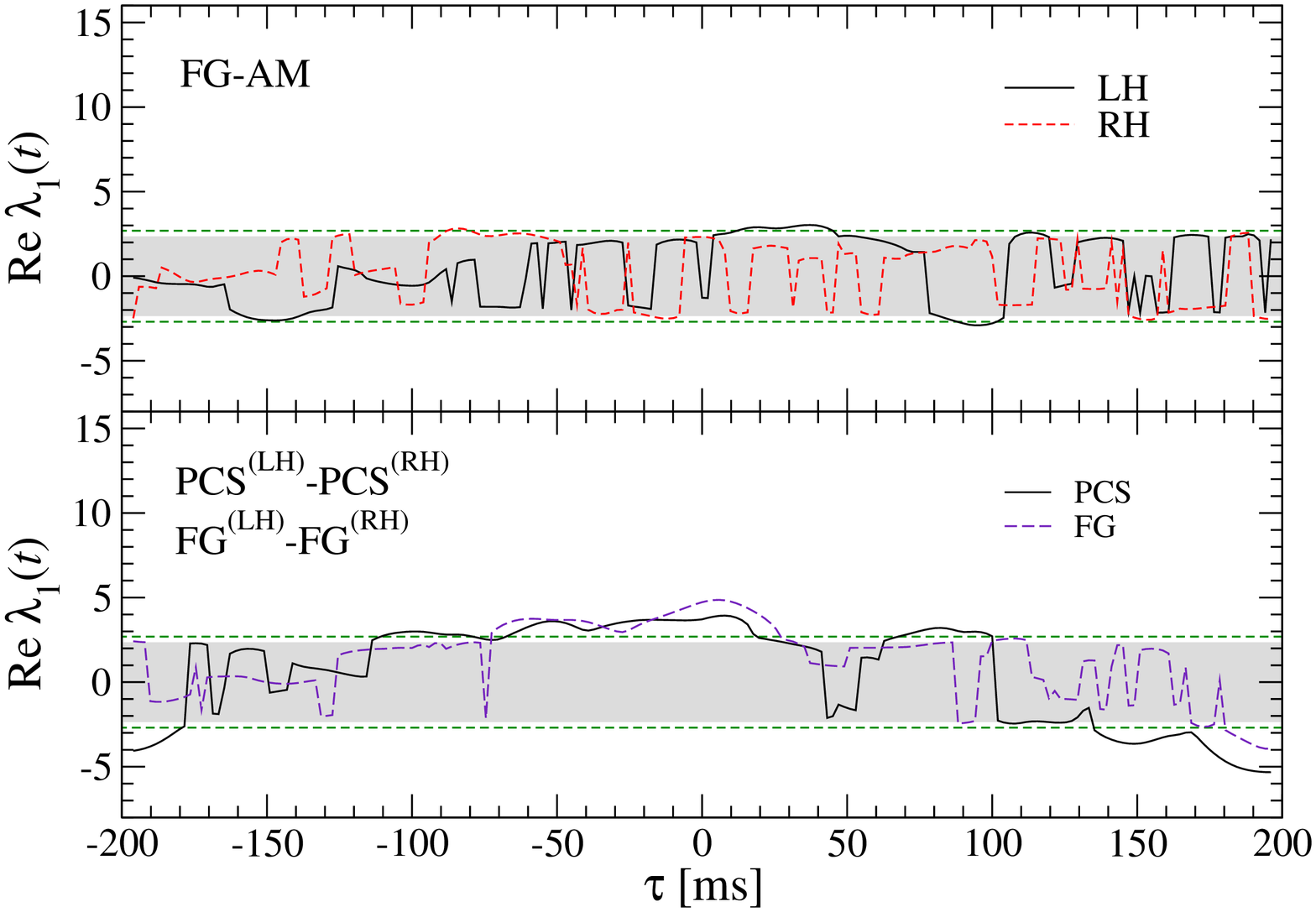} 
\caption{Real part of the largest eigenvalue $\lambda_1$ as a function of time lag $\tau$ between activity of different regions of interest located in the left (L) or in the right (R) hemisphere. $\tau$ indicates the lag of the second region in a given pair. Shaded horizontal belt denotes the asymptotic ($N \to \infty$) eigenvalue zone for the GinOE ensemble of random matrices ($|\lambda_x| \le \sqrt{N}$), while horizontal dashed lines denote value for which $\rho_G^r(\lambda) = 0.01 \rho_G^r(0)$ for $N=140$. Statistically significant non-random values of $\lambda_1(\tau)$ are purely real.} 
\label{fig::largest.eigenvalue} 
\end{figure} 

\begin{figure} 
\hspace{1.8cm} 
\epsfxsize 8cm 
\epsffile{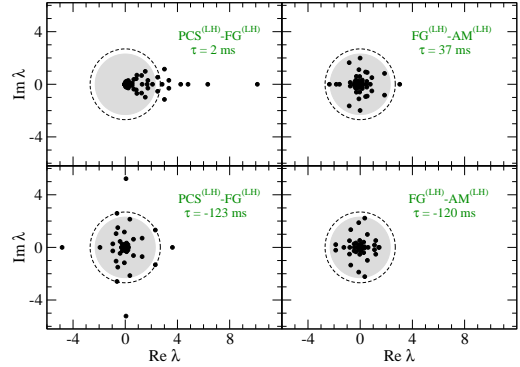} 
\caption{Exemplary eigenvalue spectra of asymmetric correlation matrix ${\bf C}(\tau)$ in the complex plane. Meaning of the shaded full circles and the dashed circles centered at (0,0) are the same as meaning of their counterpart zones in Figure~\ref{fig::largest.eigenvalue}. Most eigenvalues are concentrated around (0,0) due to eigenvalue properties of the asymmetric Wishart matrices whose spectra are radially inhomogeneous in contrast to the GinOE matrices.} 
\label{fig::eigenvalue.spectrum} 
\end{figure}

Figure~\ref{fig::eigenvalue.spectrum} exhibits the exemplary full spectra of complex eigenvalues corresponding to the two pairs of regions (left hemisphere): PCS-FG (left column) and FG-AM (right column). In both cases we chose two characteristic values of $\tau$ representing the maximum of $\lambda_1(\tau)$ and a typical value related to lack of significant correlations. A typical deviation from the GinOE case is seen in all panels, where the eigenvalues have strikingly inhomogeneous distribution tending to concentrate around the (0,0) point. 

\begin{figure} 
\hspace{1.8cm} 
\epsfxsize 8cm 
\epsffile{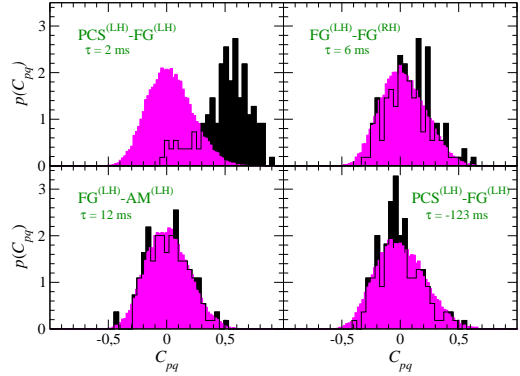} 
\caption{Distributions of diagonal (dark) and off-diagonal (light) elements of ${\bf C}(\tau)$ for different characteristic situations: large real $\lambda_1$ (highly correlated ROI activity, top left), real $\lambda_1$ slightly exceeding the GinOE threshold (weakly correlated ROI activity, top right), and two typical cases of statistically insignificant correlations with complex (GinOE) $\lambda_1$ (bottom left and right). In each case the distributions were derived either from $N$ diagonal or $N(N-1)$ off-diagonal elements, which explains smoother shapes of the distributions for the latter.}
\label{fig::matrix.elements} 
\end{figure}

Our results raise another question, whether the activation of the analyzed ROIs is repeatable, i.e., whether the corresponding patterns of the stimulus-evoked activity are similar in different trials or they are variable. This issue might be addressed owing to the fact that the same pattern of activity in signals associated with different trials can produce correlations of comparable strength no matter if one looks at the correlations between the simultaneous signals or one looks at correlations between the signals representing different trials (e.g., signal 12 from ROI$_1$ and signal 35 from ROI$_2$). If this is the case, we will obtain statistically similar distributions of matrix elements on the diagonal (extracting the timecourses of each ROI from the same trials) and off-diagonal (extracting the time courses of each ROI from different trials).  We note that this conclusion involves the implicit assumption that the spontaneous activity differs in different ROIs, which is close to reality. Figure~\ref{fig::matrix.elements} shows four exemplary distribution pairs for the diagonal and the off-diagonal elements calculated for four different ROI pairs. For at least one of the cases in this Figure there is strong evidence that these distributions differ significantly. In fact, for $\tau$ corresponding to the maximum value of $\lambda_1(\tau)$ in the left PCS-FG pair (top left panel) the typical diagonal elements are much larger than the typical off-diagonal elements. This indicates that simultaneous correlations are considerably stronger than the cross-trial ones. It comes straightforward thus that the patterns of each ROI activity may be different in consecutive trials. This can be partly accounted for by the fact that the subject was presented with the images of different objects which can also be processed differently by the ROIs. A more detailed analysis is required where the correlation matrices are constructed separately within each category of objects. This, however, exceeds the scope of the present work intended to be only an illustration of the methods. The other examples in Figure~\ref{fig::matrix.elements} clearly support the random character of correlations in the FG-AM pair and in the PCS-FG pair outside the correlation-related $\tau$ range (bottom panels) and suggests the existence of small but possibly significant differences between the distributions for the FG$^{(L)}$-FG$^{(R)}$ pair (top right panel). Interpretation of the latter results is similar to the above one related with the top left panel of Figure~\ref{fig::matrix.elements}. 

\section{Summary} 

Need for a theory of non-Hermitian ensembles of random matrices with complex eigenvalues has so far been identified in such diverse areas like random networks~\cite{Timme2004}, quantum chaos~\cite{Drozdz1996,Fyodorov1997}, quantum scattering phenomena~\cite{Drozdz2000} or quantum chromodynamics~\cite{Stephanov1996,Janik1997,Akemann2004}, among others. In the present contribution, we offer another example where real but explicitly asymmetric matrices emerge out of empirical multivariate data. The further development of the theory for asymmetric correlation matrices therefore seems to have many potential and significant applications. In principle any really complex system is at least partly driven by time-lagged correlations as the ones detected in the brain or in the financial markets. The use of the Wishart type matrices for the derivation of reference baseline distributions is therefore limited. The necessary generalization of the theory is even more subtle than for the complex non-Hermitian matrices. Here, even asymptotically, the distribution of eigenvalues on the complex plane, especially at the edge where relevant departures from randomness in the empirical correlation matrices may occur, depends on the ratio between the number of observations (degrees of freedom) and their length in time similarly as the Mar\u cenko-Pastur~\cite{Marcenko1967} distribution does for the ensembles of random symmetric correlation matrices. Deriving its counterpart for the asymmetric correlation matrices emerges as a necessary but demanding intellectual challenge.

\end{document}